\documentclass[12pt, onecolumn, draftclsnofoot]{IEEEtran}%

\usepackage[T1]{fontenc}
\usepackage{amsmath,amssymb,amsfonts,mathrsfs,bm}
\usepackage{amsthm}
\usepackage{cite}
\usepackage{array}
\usepackage[shortlabels]{enumitem}
\usepackage{graphicx}
\usepackage{url}
\usepackage{color}
\usepackage{algorithm,algorithmic}
\usepackage{multirow}
\usepackage[table]{xcolor}
\usepackage[normalem]{ulem}
\usepackage{array}
\newcolumntype{L}[1]{>{\raggedright\let\newline\\\arraybackslash\hspace{0pt}}m{#1}}
\newcolumntype{C}[1]{>{\centering\let\newline\\\arraybackslash\hspace{0pt}}m{#1}}
\newcolumntype{R}[1]{>{\raggedleft\let\newline\\\arraybackslash\hspace{0pt}}m{#1}}
\usepackage{xparse}

\ifx\useTheoremCounter\undefined
\newtheorem{Theorem}{Theorem}
\newtheorem{Corollary}{Corollary}
\newtheorem{Proposition}{Proposition}
\newtheorem{Lemma}{Lemma}
\else
\newtheorem{Theorem}{Theorem}

\newtheorem{Proposition}[Theorem]{Proposition}
\newtheorem{Lemma}[Theorem]{Lemma}
\fi

\newtheorem{Definition}{Definition}
\newtheorem{Example}{Example}
\newtheorem{Remark}{Remark}

\theoremstyle{remark}

\DeclareSymbolFont{bsfletters}{OT1}{cmss}{bx}{n}
\DeclareSymbolFont{ssfletters}{OT1}{cmss}{m}{n}
\DeclareMathSymbol{\bsfGamma}{0}{bsfletters}{'000}
\DeclareMathSymbol{\ssfGamma}{0}{ssfletters}{'000}
\DeclareMathSymbol{\bsfDelta}{0}{bsfletters}{'001}
\DeclareMathSymbol{\ssfDelta}{0}{ssfletters}{'001}
\DeclareMathSymbol{\bsfTheta}{0}{bsfletters}{'002}
\DeclareMathSymbol{\ssfTheta}{0}{ssfletters}{'002}
\DeclareMathSymbol{\bsfLambda}{0}{bsfletters}{'003}
\DeclareMathSymbol{\ssfLambda}{0}{ssfletters}{'003}
\DeclareMathSymbol{\bsfXi}{0}{bsfletters}{'004}
\DeclareMathSymbol{\ssfXi}{0}{ssfletters}{'004}
\DeclareMathSymbol{\bsfPi}{0}{bsfletters}{'005}
\DeclareMathSymbol{\ssfPi}{0}{ssfletters}{'005}
\DeclareMathSymbol{\bsfSigma}{0}{bsfletters}{'006}
\DeclareMathSymbol{\ssfSigma}{0}{ssfletters}{'006}
\DeclareMathSymbol{\bsfUpsilon}{0}{bsfletters}{'007}
\DeclareMathSymbol{\ssfUpsilon}{0}{ssfletters}{'007}
\DeclareMathSymbol{\bsfPhi}{0}{bsfletters}{'010}
\DeclareMathSymbol{\ssfPhi}{0}{ssfletters}{'010}
\DeclareMathSymbol{\bsfPsi}{0}{bsfletters}{'011}
\DeclareMathSymbol{\ssfPsi}{0}{ssfletters}{'011}
\DeclareMathSymbol{\bsfOmega}{0}{bsfletters}{'012}
\DeclareMathSymbol{\ssfOmega}{0}{ssfletters}{'012}

\DeclareMathOperator*{\argmin}{arg\,min}

\newcommand{\qednew}{\nobreak \ifvmode \relax \else
	\ifdim\lastskip<1.5em \hskip-\lastskip
	\hskip1.5em plus0em minus0.5em \fi \nobreak
	\vrule height0.75em width0.5em depth0.25em\fi}

\newcommand{\norm}[1]{{\left\lVert{#1}\right\rVert}}

\newcommand{\cond}[2]{\left. {#1}\, \middle| \, {#2} \right.}
\DeclareDocumentCommand \P { g d() g } {%
	\IfNoValueTF {#3} 
	{%
		\IfNoValueTF {#1} 
		{%
			\IfNoValueTF {#2}
			{%
				\mathbb{P}%
			}%
			{%
				\mathbb{P}\left({#2}\right)%
			}%
		}%
		{%
			\IfNoValueTF {#2}
			{%
				\mathbb{P}_{#1}%
			}%
			{%
				\mathbb{P}_{#1}\left({#2}\right)%
			}%
		}%
	}%
	{%
		\IfNoValueTF {#1} 
		{%
			\mathbb{P}\left(\cond{#2}{#3}\right)%
		}%
		{%
			\mathbb{P}_{#1}\left(\cond{#2}{#3}\right)%
		}%
	}%
}

\DeclareDocumentCommand \E { g o g } {%
	\IfNoValueTF {#3} 
	{%
		\IfNoValueTF {#1} 
		{%
			\IfNoValueTF {#2}
			{%
				\mathbb{E}%
			}%
			{%
				\mathbb{E}\left[{#2}\right]%
			}%
		}%
		{%
			\IfNoValueTF {#2}
			{%
				\mathbb{E}_{#1}%
			}%
			{%
				\mathbb{E}_{#1}\left[{#2}\right]%
			}%
		}%
	}%
	{%
		\IfNoValueTF {#1} 
		{%
			\mathbb{E}\left[\cond{#2}{#3}\right]%
		}%
		{%
			\mathbb{E}_{#1}\left[\cond{#2}{#3}\right]%
		}%
	}%
}

\definecolor{gray90}{gray}{0.9}

\newcommand{\blue}[1]{{{\color{blue} #1}}}

\newcommand{\hide}[1]{}

\graphicspath{{./Figures/}}

\usepackage{graphicx}
\usepackage{tikz}
\usepackage{color}
\usepackage{pgfplots}
\usepackage{cite}
\pgfplotsset{compat=1.5}

\providecommand{\U}[1]{\protect\rule{.1in}{.1in}}

\theoremstyle{definition}

\begin{document}
\date{}
\title{Signal Processing on Simplicial Complexes}
\author{Feng~Ji, Giacomo Kahn, and Wee~Peng~Tay,~\IEEEmembership{Senior Member,~IEEE}
\thanks{The authors are with the School of Electrical and Electronic Engineering, Nanyang Technological University, 639798, Singapore (e-mail: jifeng@ntu.edu.sg, giacomo.kahn@gmail.com, wptay@ntu.edu.sg).}}
\maketitle
\begin{abstract}
Theoretical development and applications of graph signal processing (GSP) have attracted much attention.
In classical GSP, the underlying structures are restricted in terms of dimensionality. A graph is a combinatorial object that models binary relations, and it does not directly model complex $n$-ary relations. One possible high dimensional generalization of graphs are simplicial complexes. They are a step between the constrained case of graphs and the general case of hypergraphs. In this paper, we develop a signal processing framework on simplicial complexes, such that we recover the traditional GSP theory when restricted to signals on graphs. It is worth mentioning that the framework works much more generally, though the focus of the paper is on simplicial complexes. We demonstrate how to perform signal processing with the framework using numerical examples. 
\end{abstract}


\section{Introduction}
Many fields of research use data to create hypotheses and try to infer them. Due to this heterogeneous landscape, the data themselves come in diverse forms, from simple binary relations to relations of high arity. One way of incorporating topological properties in data analysis is to use graph signal processing (GSP) \cite{Shu13, San13, San14, Don16, ort17, Gra18, Ort18, JiTay:J19}. From signals recorded on networks such as sensor networks, GSP uses graph metrics that relay the topology of the graph to perform sampling, translating and filtering of the signals. Recently, GSP based graph convolution neural network also receives much attention \cite{Def16,Kip16,ort17,Wan18}. 

\begin{figure}
	\footnotesize
	\centering
	\begin{minipage}[b]{.5\linewidth}
		\centering
		\centerline{\includegraphics[scale=0.2]{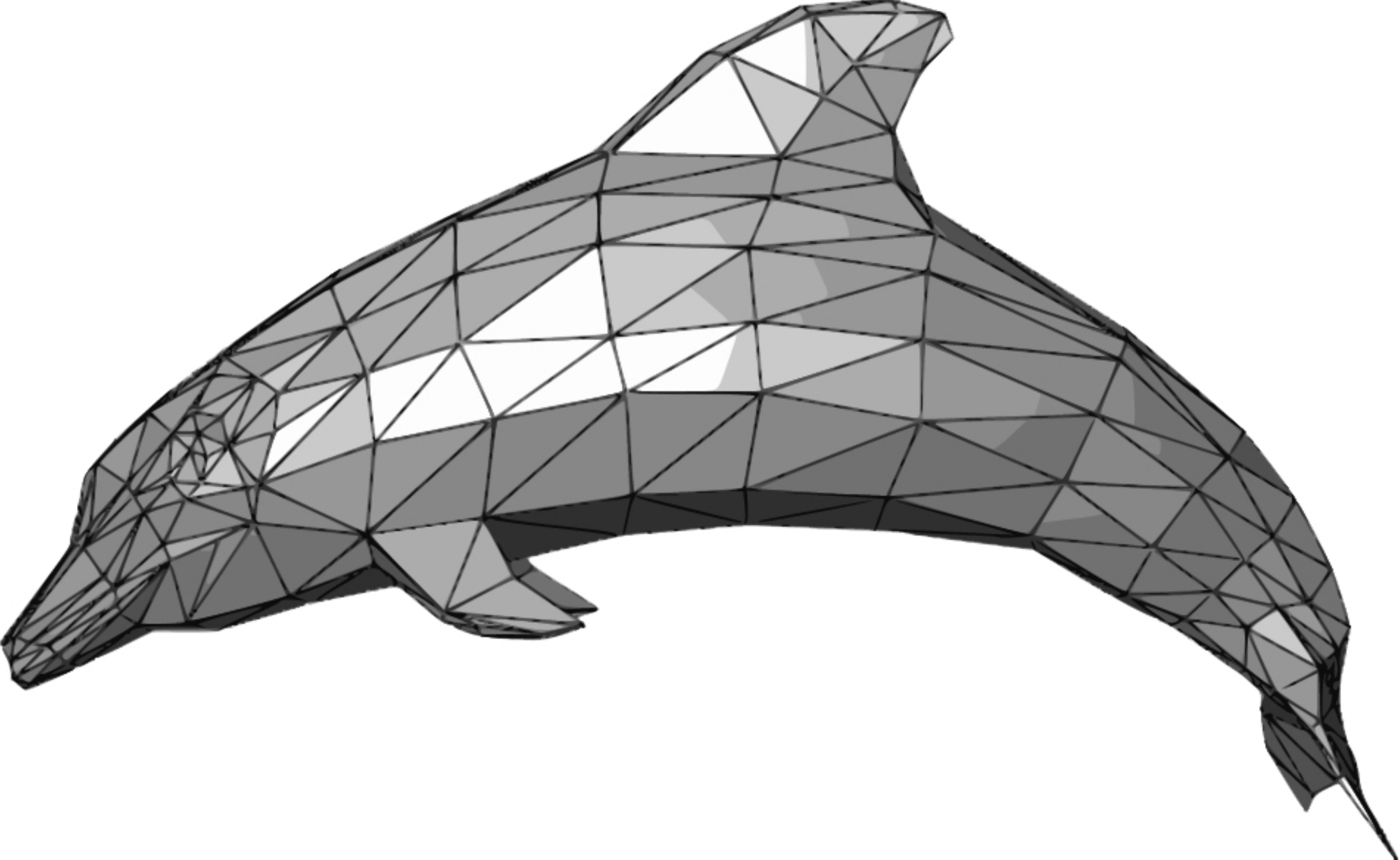}}
		\centerline{(a)}
	\end{minipage}%
	\begin{minipage}[b]{.5\linewidth}
		\centering
		
		\centerline{\includegraphics[scale=0.4]{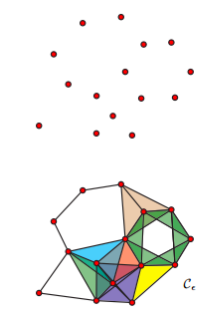}}
		\centerline{(b)}
	\end{minipage}
	\caption{(a) Approximation of a $2$D surface with a $2$-complex from Wikimedia Commons. (b) Model of a point cloud (top), together with a simplicial complex (bottom).} \label{fig:1}
\end{figure}

GSP, as useful a tool as it is, still has its limitations. The vast data landscape includes complex data, such as high dimensional manifolds, or point clouds possessing high dimensional geometric features (cf.\ Figure~\ref{fig:1}). For example, $2$D meshes can be used to approximate a surface and high dimensional simplicial complexes can be used to model discrete point clouds. Another example is in social networks such as Facebook, where an edge represents the friend relation but higher arity edges can represent family links or the inclusion in the same groups. This model also works for group conversations or other user groups in social networks. In addition, some complex interactions cannot be fully grasped by reducing them to binary relations. This is the case with chemical data, where two molecules might interact only in the presence of a third that serves as catalyst~\cite{klamt2009hypergraphs, flamm2015generalized}, or with datasets such as folksonomies, where data are ternary or quaternary relations (users$\times$ressources$\times$tag)~\cite{lohmann2012representing}. Therefore, it is then necessary to go beyond graphs to fully capture these more complex interaction mechanisms. 

Moreover, in a lot of applications, a graph learning procedure is involved based on information such as geometric distance, node feature similarity and graph signals \cite{Alt92, Don19, Mat19, Ram19, Ji20}; and due to the lack of definite meaning of edge connections, it is arguable whether a graph is the best geometric object for signal processing. For these reasons, there is a need for a framework that permits signal processing on such high dimensional geometric objects. 

Simplicial complexes, as high dimensional generalization of graphs, are independently found in many fields of computer science and mathematics. They are commonly introduced as generalizations of triangles to higher dimensions. Moreover, it is well known that simplicial complexes can be used to model and approximate any reasonable (topological) space according to the \emph{simplicial approximation theorem} \cite{Spa66}. 
They are already used in topological data analysis\cite{carlsson2009topology}, representation of surfaces in high dimensions, and modeling of complex networks~\cite{Cou16}. In this paper, we pursue a signal processing framework on simplicial complexes.  

Despite the fact that the subject is relatively new, a few attempts have been made. In \cite{bar09, bar16} the authors develop a signal processing framework using a differential operator on simplicial chain complexes. It considers signals associated with high dimensional simplices such as edges and faces, and not only vertices. The paper \cite{pus19} proposes an approach on meet or join semi-lattices that uses lattice operators as the shift and \cite{zha19} proposes a framework on hypergraphs using tensor decomposition. 

In our paper, we propose another signal processing framework for signals on nodes of simplicial complexes. Our approach makes full use of the geometric structures and strictly generalizes traditional GSP by introducing generalized Laplacians. Signal processing tasks can thus be performed similar to traditional GSP. The rest of the paper is organized as follows. We recall fundamentals of simplicial complexes in Section~\ref{sec:simp}. In Section~\ref{sec:lap}, we introduce a general way to construct Laplacian on metric spaces, and it is applied simplicial complexes. We focus on the special case of $2$-complexes in Section~\ref{sec:2cs}. In Section~\ref{sec:sp}, we describe the procedure to construct $2$-complex structures on a given graph, and discuss how to perform signal processing tasks. We present simulation results in Section~\ref{sec:sim} and conclude in Section~\ref{sec:con}. 

\section{Simplicial complexes} \label{sec:simp}
In this section, we give a brief self-contained overview of the theory of simplicial complexes (see \cite{Spa66, Hat02} for more details).  

\begin{figure}[!ht]
	\centering
	\includegraphics[scale=0.80]{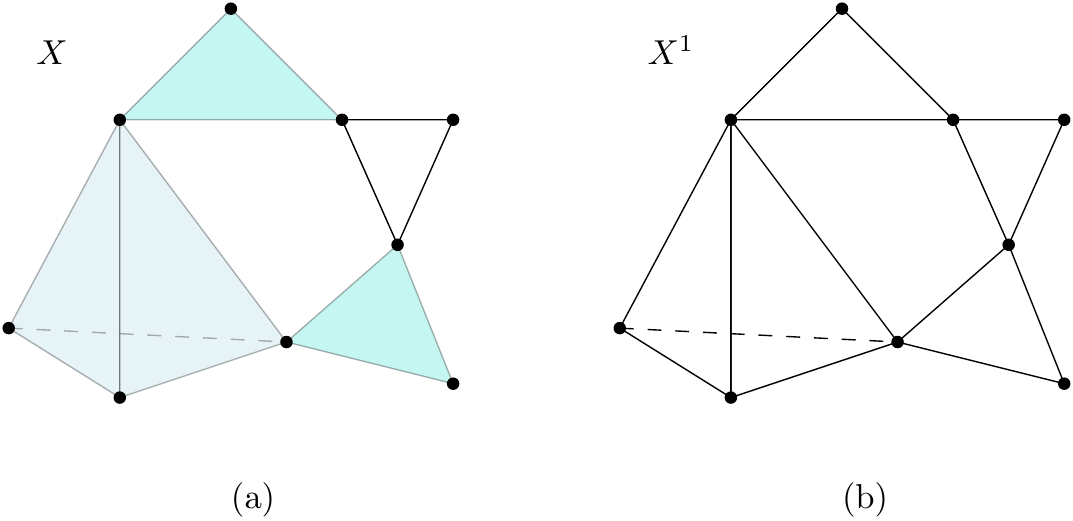}
	\caption{(a) $X$ is (the geometric realization) of a $3$-complex with a maximal $3$-simplex, $2$ maximal $2$-simplices and $3$ maximal $1$-simplices. (b) $X^1$ is a connected graph with $15$ edges.}\label{fig:sc9}
\end{figure}

\begin{Definition} \label{defn:tss}
The \emph{standard $n$-simplex (or dimension $n$ simplex) $\Delta_n$} is defined as 
\begin{align*}
\{x_0+\ldots + x_n=1\mid (x_0,\ldots, x_n) \in \mathbb{R}_+^{n+1}\}. 
\end{align*}
Any topological space homeomorphic to the standard $n$-simplex is called an $n$-simplex. In $\Delta_n$, if we require $k>0$ coordinates being $0$, we obtain an $(n-k)$-simplex, called a \emph{face}. 

A \emph{simplicial complex} $X$ (see Figure~\ref{fig:sc9} for an example) is a set of simplices such that any face from a simplex of $X$ is also in $X$ and the intersection for any two simplices $\sigma_1,\sigma_2$ of $X$ is a face of both $\sigma_1$ and $\sigma_2$. A simplex of $X$ is called \emph{maximal} if it is not the face of any other simplices. 
\end{Definition}

We shall primarily focus on finite simplicial complexes, i.e., a finite set of simplices. The dimension $\dim X$ of $X$ is the largest dimension of a simplex in $X$. For each $m\geq 0$, the subset of $m$-simplices of $X$ and their faces is called its \emph{$m$-skeleton}, denoted by $X^m$. 

Combinatorially, if we do not want to specify an exact homeomorphism of a $n$-simplex $X$ with $\Delta_n$, we may just represent it by $n+1$ labels. Therefore, its faces are just subset of the labels. It is worth pointing out that according to the above definition, a simplicial complex is a set of spaces, each homeomorphic to a simplex and they are related to each other by face relations. However, it is possible to produce a concrete geometric object for each simplicial complex.

\begin{Definition}
The \emph{geometric realization} of a simplicial complex $X$ is the topological space obtained by gluing simplices with common faces. 
\end{Definition}

\begin{Example}
Let $X$ be a finite simplicial complex consisting of two types of simplexes $E$ and $V$. Each simplex in $E$ is $1$-dimensional and $V$ contains only $0$ simplexes. The geometric realization of $X$ is nothing but a graph with vertex set $V$ and edge set $E$. More generally, for any simplicial complex $X$, the geometric realization of $X^1$ is a graph in the usual sense.
\end{Example}

For convenience, we shall not distinguish simplicial complex with its geometric realization when no confusion arises.  

In this paper, a simplicial complex $X$ is weighted if $X^1$ is a weighted graph. Otherwise, we may make an unweighted simplicial complex become weighted by assigning length $1$ to each $1$-simplex. In this way, $X^0$ becomes a metric space.

A notion even more general than ``simplicial complex" is \emph{hypergraph} \cite{Ber73, Ouv20}. A hypergraph $H=(V,E)$ is a pair where $V$ is a set of vertices and $E$ is a set of non-empty subsets of $V$, called hyperedges. Each simplicial complex $X$ might be viewed as a hypergraph $H_X=(X^0, E)$, where the vertex set is the $0$-skeleton of $X$ and vertices of each simplex of dimension at least $1$ give a hyperedge. Conversely, a hypergraph may not arise from a simplicial complex as such since a subset of a hyperedge might not be a hyperedge, violating the face condition of Definition~\ref{defn:tss}. For each hypergraph $H$, there is an associated simplicial complex $X_H$ whose simplices are spanned by hyperedges in $E$ as well as their faces. Therefore, the signal processing framework developed in this paper can be directly applied to each hypergraph $H$, or more precisely, the associated simplicial complex $X_H$.

\section{Generalized Laplacian} \label{sec:lap}

Recall that graph signal processing relies heavily on the notion of ``shift operator". A popular choice is the graph Laplacian. In this section, we want to generalize this notion.

\begin{Definition} \label{defn:lxb}
	Let $X$ be a finite metric space of size $|X|=n$. A generalized Laplacian consists of the following data:
	\begin{enumerate}[(A)]
		\item a weighted, undirected graph $G_X=(V,E)$, 
		\item a set function $f: X^0 \to V$, and
		\item a linear transformation $T: \mathbb{R}^{|X^0|} \to \mathbb{R}^{|V|}$,
	\end{enumerate}
	such that the following holds:
	\begin{enumerate}[(a), nolistsep]
		\item $f$ is one-one,
		\item \label{it:tfc} the $f(v)$ component of $T(x)$ is the same as the $v$ component of $x$ for each $v\in X^0$ and $x \in \mathbb{R}^{|X^0|}$, and
		\item \label{it:tso} the sum of each row of $T$ (written as a transformation matrix) is a constant.
	\end{enumerate}
	Let $L_{G_X}$ be the Laplacian of the weighted graph $G_X$. The \emph{generalized Laplacian associated with the data $(G_X,f,T)$} is defined as 
	\begin{align*}
	L_{(G_X,f,T)} = T'\circ L_{G_X} \circ T: \mathbb{R}^{|X^0|} \to \mathbb{R}^{|X^0|},
	\end{align*}
	where $T'$ denotes the adjoint (transpose) of $T$.  We abbreviate $L_{(G_X,f,T)}$ by $L_X$ if no confusion arises from the context.
\end{Definition}

Intuitively, we require that $f$ is one-one to ensure that $f$ ``embeds" $X$ in $G_X$ such that we may perform the shift operation on $G_X$. Conditions~\ref{it:tfc} and \ref{it:tso} on $T$ say that the signal on $v\in X^0$ is preserved at its image $f(v)$ in $G_X$, while signals on $V\backslash f(X)$ are formed from an averaging process.

\begin{Lemma} \label{lem:llc}
	\begin{enumerate}[(a)]
		\item $L_X$ is symmetric.
		
		\item $L_X$ is positive semi-definite.
		
		\item Constant signals are in the $0$ eigenspace of $L_X$. The $0$-eigenspace $E_0$ of $L_X$ is $1$-dimensional if and only if $G$ is connected.
	\end{enumerate}
\end{Lemma}

\begin{IEEEproof}
	(a) $L_X$ is symmetric because $G_X$ is assumed to be undirected and hence $L_G$ is symmetric.
	
	(b) Similar to (a), $L_X$ is positive semi-definite because $L_{G_X}$ is positive semi-definite.
	
	(c) As we assume that the sum of each row of $T$ is a constant, therefore if $x$ is a constant vector, then so is $T(x)$. Since constant vectors are in the $0$-eigenspace of $L_{G_X}$, we have $L_{G_X}\circ T(x)=0$ and so is $L_X(x)$. Therefore, the dimension of $E_0$ is at least $1$.
	
	Now assume that $x$ is in $E_0$. Therefore, 
	\begin{align*}
	0 = \langle x, L_X(x) \rangle = \langle T(x), L_{G_X}\circ T(x)\rangle.
	\end{align*}
	Consequently, $T(x)$ belongs to the $0$-eigenspace of $L_{G_X}$, which is $1$-dimensional if and only if $L_G$ is connected.
	
	By Condition~\ref{it:tfc}, the operator $T$ is injective. Therefore, $E_0$ is $1$-dimensional if and only if $T(E_0)$ is $1$-dimensional, which is in turn equivalent to $G_X$ being connected as we just observe.
\end{IEEEproof}

By Lemma~\ref{lem:llc}, the generalized Laplacian $L_X$ enjoys a few desired properties. In particular, being symmetric permits an orthonormal basis consisting of eigenvectors of $L_X$. Therefore, one can devise a Fourier theory analogous to traditional GSP. Moreover, as $L_X$ is positive semi-definite, we may perform smoothness based learning. The constant vectors belong to the $0$-eigenspace is also desirable as it agrees with the intuition that ``constant signals are smoothest". 

On the other hand, Lemma~\ref{lem:llc} asserts that $L_X$ is indeed very similar to the Laplacian of a graph. The theory will be less useful if we are only able to produce weighted graph Laplacian, which we shall prove to be untrue. To this point, we introduce the following notion.

\begin{Definition}
We call $L_X$ \emph{graph type} if all the diagonal entries of $L_{X}$ are non-negative and all the off-diagonal entries are non-positive.
\end{Definition}

Now for simplicial complexes, we are going to give an explicit construction of $L_X$ together with the choice of $G, f$ and $T$. As an implicit requirement, we would like the construction to recover the usual Laplacian if $X$ is a graph. 

For the simplest case, assume $X \cong \Delta_n$ is a weighted $n$-simplex, i.e., $X$ is homeomorphic to the standard $n$-simplex and its $1$-skeleton $X^1$ is a weighted graph. We label the vertices of $X$ by $v_1,\ldots, v_{n+1}$. The graph $G_X=(V,E)$ is constructed as follows: $V = \{v_1,\ldots, v_{n+1},u\}$ with a single additional vertex $u$, which is understood as the barycenter of $X$. There is no edge between $v_i$ and $v_j$ for any pair $1\leq i\neq j \leq n+1$. On the other hand, there is an edge connecting $v_i$ and $u$ for each $1\leq i\leq n+1$.

\begin{figure}[!ht]
	\centering
	\includegraphics[scale=1.2]{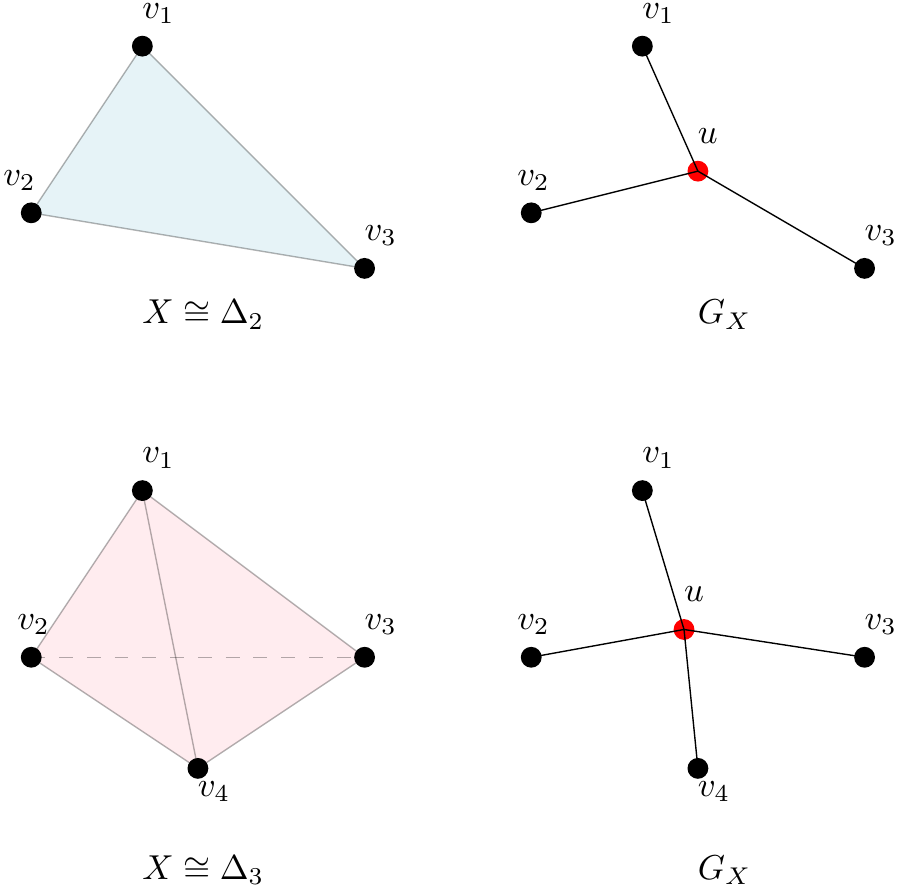}
	\caption{Graphical illustration of the shape of $G_X$ for $X \cong \Delta_2$ and $X \cong \Delta_3$.}\label{fig:sc1}
\end{figure}

The edge weight $w(v_i,u)$ between $v_i$ and $u$ is computed as follows:
\begin{align} \label{eq:wvu}
w(v_i,u) = \frac{1}{\binom{n}{2}} (\sum_{v_i\neq v_j\neq v_k\neq v_i} (v_j,v_k)_{v_i}),
\end{align}
where 
\begin{align*}
(v_j,v_k)_{v_i} =(d_X(v_i,v_j)+d_X(v_i,v_k)-d_X(v_j,v_k))/2,
\end{align*}
the Gromov product \cite{Kap02, JiT19}. Illustrations for $X \cong \Delta_2$ and $X \cong \Delta_3$ are shown in Figure~\ref{fig:sc1}. When $n=2$, in $G_X$, we recover pairwise distances between the nodes $v_1, v_2$ and $v_3$ in $X$. 

We have a canonical choice for $T$: $T(v_i)=v_i$. For $f$, it is identity on each $v_i$ component, while the average is assigned to the $u$ component. In the matrix form, 
\begin{align*}
f = \begin{bmatrix}
1 & 0 & \ldots  & 0 \\
0 & 1 & \ldots  & 0 \\
\vdots & \vdots & \ddots & \vdots \\
0 & 0 &\ldots & 1 \\
1/(n+1) & 1/(n+1) & \ldots & 1/(n+1)
\end{bmatrix}.
\end{align*} 

It is straightforward to check that $(G_X,T,f)$ verifies the conditions of Definition~\ref{defn:lxb}. Thus, we have an associated generalized Laplacian $L_X$. 

For a general finite simplicial complex $X$, we have a decomposition $X^{max}$ as the subset of the maximal simplices in $X$ and the generalized Laplacian 
\begin{align*}
L_X =  \sum_{\sigma \in X^{max}} L_{\sigma}, 
\end{align*} 
where the summation is over all maximal simplices of $X$ with appropriate embedding of the vertex indices of $\sigma$ in $X$.

To give some insights of the construction, we notice that for $X \cong \Delta_n$, it is topologically (homotopy) equivalent to a point \cite{Hat02}. Therefore, if we want to approximate $X$ by a graph $G_X$ that preserves this topological property, $G_X$ must be a tree. In addition, if we do not want to break the symmetry of the vertices, the most natural step to do so is to add one additional node (the barycenter) connected to every vertex in the original graph. The edge weights of $G_X$ are chosen to approximate the metric of $X$. 

Other evidence for the construction shall be discussed in subsequent sections. We end this section by a discussion when $X$ itself is a graph. In this case, the maximal simplexes are just edges. However, for an edge $e = (v_1,v_2)$ with weight $w$, the associated graph $G_e$ contains only $3$ nodes, $v_1,v_2$ and an additional node $u$. Therefore, in this case, Formula~(\ref{eq:wvu}) no longer applies. On the other hand, to apply Definition~\ref{defn:lxb}, we may choose $G_X$ to be $X$ itself and both $f$ and $T$ be the identify map. Therefore, we recover $L_X$ as the usual graph Laplacian.

\section{$2$-complexes} \label{sec:2cs}

In this section, we focus on $2$-complexes, on which the main applications is based on. For a weighted $2$-simplex $X \cong \Delta_2$, assume that the edge weights are $w(v_1,v_2), w(v_1,v_3)$ and $w(v_2,v_3)$. The edge weights of $G_X$ are 
\begin{align*}
& a = (v_2,v_3)_{v_1} = (w(v_1,v_3)+w(v_1,v_2)-w(v_2,v_3))/2,\\
& b = (v_1,v_3)_{v_2} = (w(v_2,v_3)+w(v_1,v_2)-w(v_1,v_3))/2,\\
& c = (v_1,v_2)_{v_3} = (w(v_1,v_3)+w(v_2,v_3)-w(v_1,v_2))/2.
\end{align*}

If the edge weights satisfy the triangle inequality, then $a\geq 0, b\geq 0, c\geq 0$. Conversely, given $a\geq 0, b\geq 0, c\geq 0$, we are able to recover the edge weights by taking pairwise sums.

The generalized Laplacian $L_X$ is thus given by:
\begin{align*}
L_{X} &= \begin{bmatrix}
1 & 0 & 0 & 1/3 \\
0 & 1 & 0 & 1/3 \\
0 & 0 & 1 & 1/3
\end{bmatrix}
\begin{bmatrix}
a & 0  & 0 & -a \\
0 & b  & 0 & -b \\
0 & 0  & c & -c \\
-a & -b  & -c & a+b+c
\end{bmatrix}
\begin{bmatrix}
1 & 0 & 0  \\
0 & 1 & 0  \\ 
0 & 0 & 1  \\
1/3 & 1/3 & 1/3
\end{bmatrix}
\\ &= \frac{1}{9}\begin{bmatrix}
b+c+4a & c-2a-2b & b-2a-2c  \\
c-2a-2b & a+c+4b & a-2b-2c  \\ 
b-2a-2c & a-2b-2c & a+b+4c  
\end{bmatrix}.
\end{align*}

\begin{Definition}
Define the \emph{shape constant} $\gamma_X$ of $X$ as 
\begin{align*}
\gamma_X = \min\{\frac{5w(v_i,v_j)-w(v_i,v_k)-w(v_j,v_k)}{2}, \{i,j,k\}=\{1,2,3\}\}.
\end{align*}
\end{Definition}

In general, $\gamma_X$ can be negative. This happens when there is at least one very short edge. We use it to address an issue left over from the previous section.

\begin{Lemma}
	Suppose $X \cong \Delta_2$ is a $2$-simplex. Then $L_X$ is of graph type if and only if $\gamma_X\geq 0$. 
\end{Lemma}

\begin{IEEEproof}
	A direct computation shows that 
	\begin{align*}
	-\gamma_X = \max\{c-2a-2b, b-2a-2c, a-2b-2c\}.
	\end{align*}
	As the diagonal entries of $L_X$ are all positive, it is of graph type if and only if $-\gamma_X \leq 0$, i.e., $\gamma_X\geq 0$.
\end{IEEEproof}

In the case of $2$-simplex, we may also give the following interpretation of $L_X$ with the graph Laplacian $L_{X^1}$ of the $1$-skeleton $X^1$. 

Consider a graph signal $x = (x_1,x_2,x_3)'$ on the vertices $\{v_1,v_2,v_3\}$. Let $y$ be the first order difference $(x_3-x_2, x_1-x_3, x_2-x_1)'$. By a direct computation, one observes that $L_X$ is determined by 
\begin{align*}
9\langle x, L_X(x)\rangle = \langle y, L_{X^1}(y) \rangle.
\end{align*}
It says $L_X$ is a higher order difference, though the point-of-view cannot be generalized beyond dimension $2$.

If $X$ is a general $2$-dimensional simplicial complex, the Laplacian $L_X$ takes contribution form Laplacian of $2$-simplexes computed as above and usual edge Laplacians. We next study spectral properties of $L_X$. In particular, we want to compare $L_X$ and $L_{X^1}$ as the latter is well-studied. Recall that $A \preceq B$ if $B-A$ is positive semi-definite.

\begin{Lemma} \label{lem:sxi}
	Suppose $X$ is a finite $2$-dimensional simplicial complex with each edge of length $1$. Let $k_{\max}$ and $k_{\min}$ be the largest and smallest numbers of $2$-simplexes that share a single edge. Then 
	\begin{align*}
	\max\{\frac{k_{\min}}{3},\frac{1}{3}\}\cdot L_{X^1} \preceq L_X \preceq \frac{k_{\max}}{3} L_{X^1}.
    \end{align*} 
\end{Lemma}

\begin{IEEEproof}
	We sketch the main idea of the proof. It suffices to show $L = L_X - \max\{\frac{k_{\min}}{3},\frac{1}{3}\}\cdot L_{X^1}$ or $L =  \frac{k_{\max}}{3} L_{X^1} - L_X$ is the Laplacian of a (possibly disconnected) graph. For this, one only needs to compute the off-diagonal entries of $L$ and show they are non-positive, which follows from direct computation. 
\end{IEEEproof}

If $X$ is a $2$D-mesh (triangulation) of a compact $2$-manifold, then $k_{\min}=1$ and $k_{\max}=2$. This is because at most two $2$-simplexes can share a common edge and along the boundary each edge is contained in a single $2$-simplex.

\begin{figure}[!ht]
	\centering
	\includegraphics[scale=1.5]{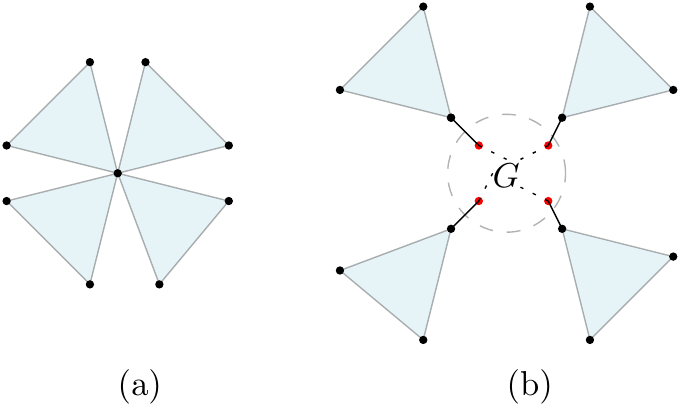}
	\caption{In (a), if all the edge weights are the same then $L_X=1/3L_{X^1}$ is shift invariant w.r.t.\ $L_{X^1}$. However, in (b), as long as the $4$ red nodes are contained in a graph $G$ (at the center), then $L_X$ is not shift invariant w.r.t.\ $L_{X^1}$ by Proposition~\ref{prop:sxi}, even if we allow arbitrary positive edge weights.}\label{fig:sc3}
\end{figure}

Recall that a filter $F$ is \emph{shift invariant} w.r.t.\ $L_{X^1}$ if $F\circ L_{X^1} = L_{X^1}\circ F$. If the graph Laplacian $L_{X^1}$ does not have repeated eigenvalues, then $F$ is shift invariant if and only if $F = P(L_{X^1})$ for some polynomial $P$ of degree at most $n-1$. The shift invariant family is of particular interest and they are readily estimated as one only has to learn the polynomial coefficients. Due to this fact, $L_X$ will be less interesting if it is shift invariant w.r.t.\ $L_{X^1}$, e.g., when $X$ is a single $2$-simplex with equal edge weights (more examples are shown in Figure~\ref{fig:sc3}). However, this does not happen in general.

\begin{figure}[!ht]
	\centering
	\includegraphics[scale=1.2]{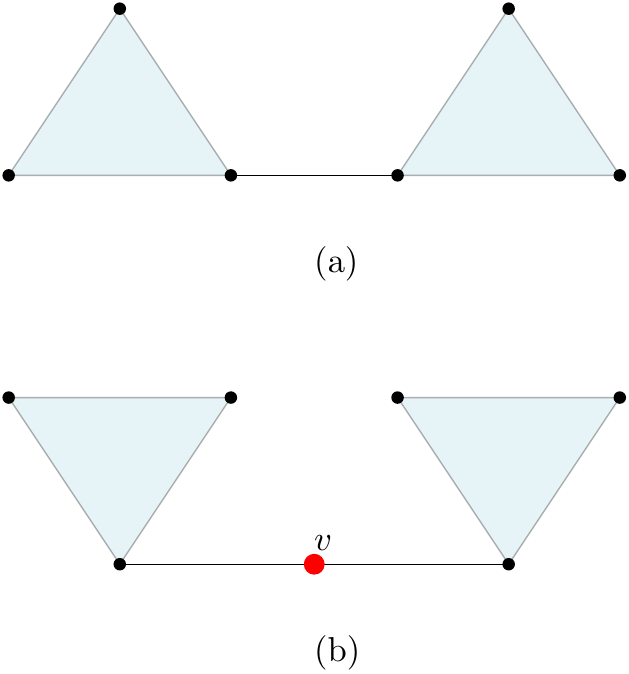}
	\caption{Illustration of the two situations disallowed by the first two conditions of Proposition~\ref{prop:sxi}.}\label{fig:sc2}
\end{figure}

\begin{Proposition} \label{prop:sxi}
Suppose $X$ is a $2$-complex such that the following condition hold (illustrated in Figure~\ref{fig:sc2}):
\begin{enumerate}[(a)]
	\item \label{it:ixa} In $X$, any two $2$-simplexes are not connected by a direct edge.
	\item \label{it:ixi} In $X$, if a vertex $v$ is not contained in any $2$-simplex, then it is connected to at most one $2$-simplex. There is at least one such vertex.
	\item \label{it:edi} Each edge is contained in at most one $2$-simplex.
\end{enumerate}
Then $L_X$ is not shift invariant w.r.t.\ $L_{X^1}$.
\end{Proposition}

\begin{IEEEproof}
	The proof is given in Appendix~\ref{sec:nsi}.
\end{IEEEproof}

\section{Learning $2$-complex structure and signal processing} \label{sec:sp}

\subsection{A family of Laplacians}

In this section, we discuss the approach to learn a $2$-complex structure $X$ given a graph $G=(V,E)$ such that $X^1=G$ and $X^0=V$ of size $n$.

For preparation, if $G$ is an unweighted graph, we assign weight $1$ to each edge. Otherwise, if pairwise similarities of $V$ are given, then we define the weight between $(v_1,v_2)$ to be the inverse of the similarity (i.e., we want two nodes to be closer if they are more similar). Therefore, we may assume that $X^1 = G$ is weighted.

The general idea goes as follows. We first identify the set $C_{X^0}$ of all possible $2$-simplexes. Depending on the problem, there are two main cases:
\begin{enumerate}[(a)]
	\item If $X^1=G$ is given, then a triple of nodes $(v_1,v_2,v_3)$ belongs to $C_{X^0}$ if and only if $(v_1, v_2), (v_1,v_3)$ and $(v_2, v_3)$ are all edges of $G$.
	
	\item If only $X^0=V$ is given, then we assume $C_{X^0}$ contains any triple $(v_1,v_2,v_3)$ of distinct nodes in $V$. 
\end{enumerate} 

Given two non-negative numbers $r_1\leq r_2$, we define $C_{X^0}(r_1,r_2)$ to be the subset of $C_{X^0}$ consisting of triples $(v_1,v_2,v_3)$ whose pairwise edge weights are within the interval $[r_1,r_2]$. Hence, we have the fundamental filtration $\emptyset = C_{X^0}(0,0) \subset C_{X^0}(0,r) \subset C_{X^0}(0,r') \subset C_{X^0}(0,\infty) = C_{X^0}$ for $r\leq r'$. Next, we perform the following steps:
\begin{enumerate}[(a),nolistsep] \label{step:oat}
	\item Order all the $2$-simplices of $C_{X^0}$ in a queue $Q$:
	\begin{enumerate}[(i),nolistsep]
		\item Choose $r_0=0 \leq r_1 \leq \ldots \leq r_m$ such that $C_{X^0} = C_{X^0}(0,r_m)$. A simplex in $C_{X^0}(0,r_{i})$ is ordered before that in $C_{X^0}(0,r_{i+1})\backslash C_{X^0}(0,r_{i})$, i.e., small triangles first.
		\item \label{it:wot} We order the $2$-simplices of $C_{X^0}(r_i,r_{i+1})$ in such a way that $2$-simplices sharing more edges are ordered later in the queue (with more details given below).
	\end{enumerate}
	\item Partition $Q$ as a disjoint union $Q = \bigcup_{1\leq i\leq p} Q_i$ such that their sizes are approximately uniform.
	\item \label{step:lxb} Let $X_0$ be $X^0\cup X^1$. For each $1\leq i \leq p$, we construct a 2-complex $X_i$ by adding the $2$-simplices of $Q_i$ (and the associated edges) to $X_{i-1}$. We form the associated generalized Laplacians $L_{X_i}$.
	\item Approximate the actual Laplacian by using one of $L_{X_i}$. This step is problem dependent, which in particular relies on the given signal and usually involves an optimization step. We shall be more explicit in Section~\ref{sec:gl}.
\end{enumerate}

\begin{figure}[!ht]
	\centering
	\includegraphics[scale=1.6]{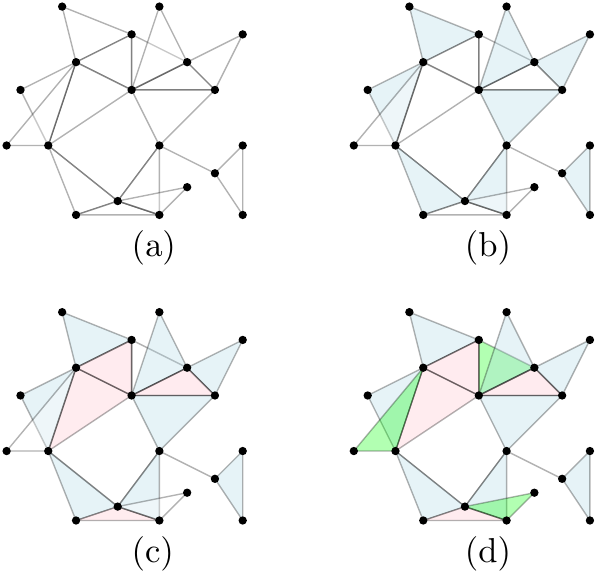}
	\caption{In this example, the blue $2$-simplexes in (b) are (randomly) ordered first in $Q$. After which, we have the pink $2$-simplexes in (c). Finally, the green $2$-simplexes are ordered last in $Q$.}\label{fig:sc5}
\end{figure}

For completeness, we describe the algorithm for Step~\ref{it:wot} (illustrated in Figure~\ref{fig:sc5}):
\begin{enumerate}[(1)]
	\item \label{it:fei} For each $i$, we randomly order the $2$-simplexes of $Q_i$ to form $Q$.
	
	\item We want to inductively re-order the members of $Q$ from the initial ordering in Step~\ref{it:fei}. We start with the first element $x_1$ of $Q$. Suppose, for $j$ ranges from the $2$-simplexes of $Q$, we have already ordered $x_1,\ldots, x_j$. Search for the rest of the $2$-simplexes of $Q$. If $x$ is sharing a common edge with $x_j$, re-order $Q$ by placing $x$ at the end of $Q$. Once all $x \in Q\backslash \{x_1,\ldots, x_j\}$ is gone through once, repeat the procedure for $x_{j+1}$.
\end{enumerate}

\subsection{Signal processing}

With any of $L$ being one of the generalized Laplacian, one may perform signal processing tasks, such as defining Fourier transform, frequency domain, sampling and filtering, similar to traditional GSP \cite{Shu13}, as we now briefly recall:

\begin{enumerate}[(a)]
	\item Fourier transform: Let $S(V)$ be the space of signals on $V$ and $\{x_i, 1\leq i\leq n\}$ be an eigenbasis of $L$. For a signal $x \in S(V)$, its \emph{Fourier transform} is given by 
	\begin{align*}
	\hat{x}(i) = \langle x,x_i \rangle, 1\leq i\leq n.
	\end{align*}
	The inverse transformation is given by:
	\begin{align*}
	x = \sum_{1\leq i\leq n} \hat{x}(i) x_i.
	\end{align*}
	
	\item Bandlimit and bandpass filters: suppose $B$ is a subset of $\{1,\ldots, n\}$. A signal $x$ has \emph{bandlimit $B$} if $\hat{x}(i)=\langle x,x_i\rangle=0$ for $i\notin B$. The bandpass filter associated with $B$ is given by $x \mapsto \sum_{i\in B} \hat{x}(i) x_i$. For denoisying  and data-compression, one may consider bandpass filters associated with $B$ consisting of small indices; while for anomaly detection, one may instead choose $B$ containing large indices. 
	
	\item Downsampling: if a signal $x$ is bandlimited with $B$ of small size, we can always have full knowledge of $x$ by looking at the signal values at a subset $V_1\subset V$ of size $|B|$. This is called \emph{downsampling}.
	
	\item Convolution: convolution is a generalization of bandpass filters. A convolution kernel is a signal $z \in S(V)$. The associated convolution filter $x\mapsto z*x$ is defined by requiring $\widehat{z*x} = \hat{z}\hat{x}$.	
	
	\item Shift invariant filters: as we have mentioned earlier, a filter $F$ is shift invariant w.r.t.\ $L$ if $F\circ L = L\circ F$. If $L$ does not have repeated eigenvalues, a shift invariant filter $F$ is always a polynomial of $L$.
\end{enumerate} 

\begin{Remark}
Though we do not use it in the paper, it is worth mentioning a continuous filter learning scheme. The basic form of the problem is specified as follows: there are two signals $x_1,x_2$ on $X^0$. Learn the structure of $X$ and an appropriate filter $F$ such that $x_2 = F(x_1)+ y$, where $y$ is the white noise. 

For $1\leq t\leq 1$, let $L_{i,t} = tL_{X_i}+(1-t)L_{X_{i+1}}$. We propose to solve the following optimization problem:
\begin{align} \label{eq:mm}
 \min_{1\leq i\leq p} \min_{\substack{t\in [0,1] \\ (a_0,\ldots, a_{b}) \in \mathbb{R}^{b+1}}} \norm{(\sum_{1\leq j\leq b} a_jL_{i,t}^j)(x_1)-x_2}^2,
\end{align}
where $b$ is a pre-determined bound on the degree of the polynomial. Doing so allows us to get access to the optimal shift $L_{i,t}$ as well as the filter $F$.
\end{Remark}

\section{Simulation results} \label{sec:sim}

\subsection{Graph learning} \label{sec:gl}

In this section, we consider three signal processing tasks: \emph{topology inference, signal compression and anomaly detection} simultaneously with synthetic experiments.

We start with the Enron email graph $G$ of size $n=500$ and $6815$ pair-wisely connected triples \cite{Kli04}\footnote{https://snap.stanford.edu/data/email-Enron.html}. We construct a $2$-complex $X$ by randomly adding $2$-simplices for pair-wisely connected triples in $G$. As a result, $G=X^1$ is observed and $X$ is unobserved. Let $B = \{f_1, \ldots, f_n\}$ be an eigenbasis of $L_X$ arranged according to increasing order of the associated eigenvalues. We randomly generate a set $S_1$ of signals from the span of the first $r_1\%$ of base signals. 

To learn $X$ from $S_1$, we construct $X_i$ and $L_{X_i}$ as in Section~\ref{sec:2cs} Step~\ref{step:lxb} for $0\leq i\leq p=20$ where $X_0=X^1=G$. Let $V_{r_1,i}$ be the matrix whose columns are the first $r_1\%$ of the eigenvalues of $L_{X_i}$. Then the estimated simplicial complex $X_b$ and its Laplacian $L_{X_b}$ is obtained by solving the optimization problem:
\begin{align*}
b = \argmin_{0\leq i\leq p=20} \sum_{f\in S_1} \norm{V_{r_1,i}V_{r_1,i}'f-f}_2^2.  
\end{align*}  

\begin{figure}[!ht]
	\centering
	\includegraphics[height=80mm, width=120mm]{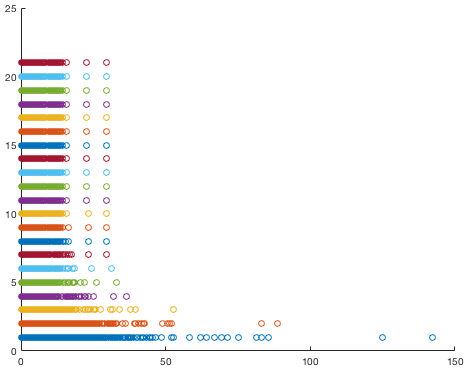}
	\caption{The plot shows the eigenvalue distribution of $L_{X_0} = L_G$ to $L_{X_{20}}$ from bottom-to-top, for the Enron email graph. The spectrum tends to shift to the left, which might show a ``more connected'' structure.}\label{fig:spec}
\end{figure}

The spectrum of $L_{X_i}$ are plotted in Figure~\ref{fig:spec}. In our case, the number of pair-wisely connected triples ($6815$) is small as compared to the size of the graph ($500$), as the number of ways choosing $3$ nodes among $n$ nodes is of order $O(n^3)$. As a consequence, the triples for our $G$ do not share many common edges. Therefore, as $i$ grows, entries of $L_{X_i}$ tend to have smaller magnitude thus yielding a shift of the spectrum towards $0$, as we observed in Figure~\ref{fig:spec} (c.f.\ Lemma~\ref{lem:sxi}). However, the situation can reverse if a graph is densely connected with a large amount of pair-wisely connected triples. Moreover, we observe that the spectrum pattern more or less stabilize beyond $i=10$, which suggests that we may choose smaller $p$ as the spectrum stratify the eigenbasis according to smoothness. We make similar observations for other non-dense graphs, e.g.\ Figure~\ref{fig:sc6}. 

Now we turn on to performance evaluation. We generate a set $S_2$ from the first $r_2\% (\leq  r_1\%)$ of the base signals in $B$, considered as a set of \emph{compressible signals}. We want to estimate the signal compression error of the estimated Laplacian $L_{X_b}$ as:
\begin{align*}
\text{err} =  \sum_{f\in S_2} \norm{V_{r_2,b}V_{r_2,b}'f-f}_2.
\end{align*}

For comparison, we perform the same estimation on $L_{X_0}$, for which we do not consider high dimensional structures. On average for different choices of $X$, as compared to using $L_{X_0}$, the compression error with $L_{X_b}$ is reduced by $33.2\%$ and $40.6\%$ for $r_1\%=r_2\%=30\%$ and $50\%$, respectively.  

\begin{figure}[!ht]
	\centering
	\includegraphics[height=75mm, width=80mm]{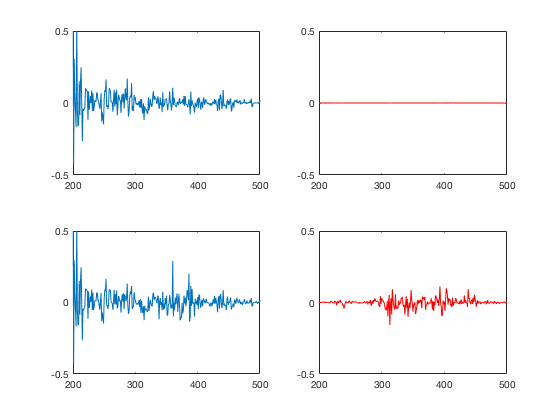}
	\includegraphics[height=75mm, width=80mm]{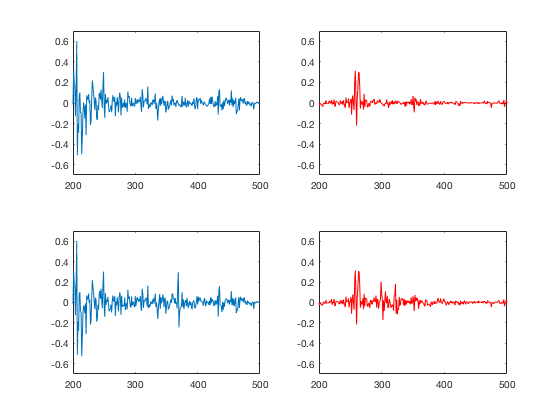}
	\caption{We have two sets of high frequency component plots. For each set, the left figures (blue) are normal and abnormal plots of $L_{X_0}$ and the right figures (red) are the plots of $L_{X_b}$. The anomalous behavior is more visible for $L_{X_b}$.}\label{fig:ano}
\end{figure}

Finally, we introduce anomalies to signals in a new set $S_3$ (with $r_1\%=2r_2\%=50\%$) by perturbing the signal value at one random node. We perform spectrum analysis of the anomalous signals using both $L_{X_b}$ and $L_{X_0}$, again as a comparison between with or without high dimensional structures. Two typical examples of the spectral plots are shown in Figure~\ref{fig:ano}. We see that using $L_{X_b}$ (red), the anomalous behavior is more easily detected by inspecting high frequency potions. 

In the subsequent subsections, we shall make more detailed investigation and analysis on anomaly detection and noisy label correction with real datasets.

\subsection{Anomaly detection} \label{sec:ano}

The graphs being used in this subsection is a weather station network in US of size $197$\footnote{http://www.ncdc.noaa.gov/data-access/land-based-station-data/station-metadata}. The dataset allows us to identify the locations of the weather stations, and the graph $G$ for the weather station network is constructed using the $k$-nearest neighbor algorithm based on the locations of the stations. The graph $G$ has $495$ edges and $395$ pair-wisely connected triples.  

\begin{figure}[!ht]
	\centering
	\includegraphics[height=120mm, width=120mm]{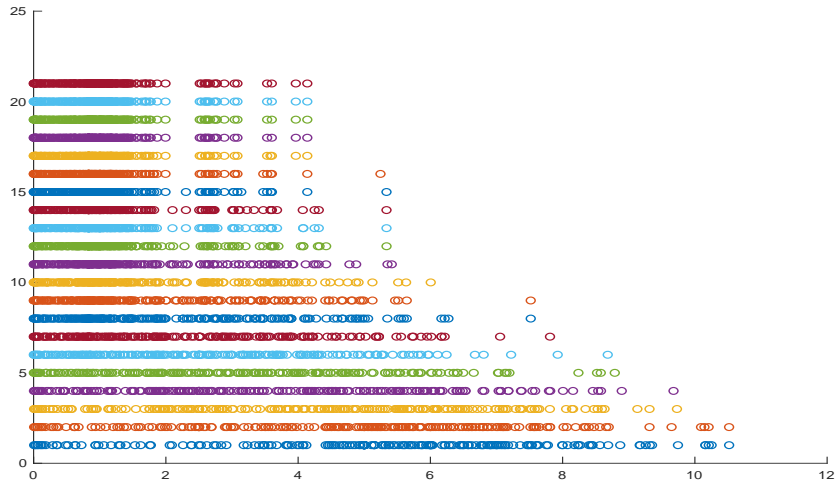}
	\caption{The plot shows the eigenvalue distribution of $L_{X_0} = L_G$ to $L_{X_{20}}$ from bottom-to-top, for a US weather station network. The spectrum tends to shift to the left, which might show a ``more connected'' structure.}\label{fig:sc6}
\end{figure}

As in Section~\ref{sec:gl}, we construct $X_i$ and $L_{X_i}$ for $0\leq i\leq p=20$. The spectrum of $L_{X_i}$ are plotted in Figure~\ref{fig:sc6} with observations similar to those made in Figure~\ref{fig:spec}.

The signals on $G_2$ are real daily temperature recorded over the year 2013\footnote{ftp://ftp.ncdc.noaa.gov/pub/data/gsod}. For a daily temperature reading $x$ chosen from the one year recording, we introduce anomaly to $x$ by randomly perturbing the value of $x$ at a single node. The resulting signal is denoted by $x_a$. 

As we mentioned in Section~\ref{sec:gl}, we may look at the high frequency components of the Fourier transform of $x_a$, decomposed w.r.t.\ $L_{X_i}$ for $0\leq i\leq p$. 

The experiment details are given as follows. We fix numbers $0<r,\epsilon<1$ and let $L$ be among $L_{X_i}$. For each instance, we randomly choose a date, and let the temperature signals on the $4$ consecutive days starting from the chosen date be $x_1, x_2, x_3$ and $x$. The signal $x_a$ is the perturbed version of $x$. Assume that we only observe the normal signals $x_1, x_2, x_3$ and the anomalous signal $x_a$. We perform graph Fourier transform on $x_1, x_2, x_3$ and $x_a$ to obtain $\hat{x}_1, \hat{x}_2, \hat{x}_3$ and $\hat{x}_a$. Define $$a = \max_{j=1,2,3; 197r<k\leq 197} |\hat{x}_j(k)|; b = \max_{197r<k\leq 197} |\hat{x}_a(k)|$$ as a measure of magnitude of the high frequency components of the signals. We declare that $x_a$ is abnormal if $b/a > 1+\epsilon$.

We run experiments with $r = 0.8$ and $\epsilon = 0.05$. We are interested in the performance in terms of percentage of successful detections under the following circumstances: 
\begin{enumerate}[S1.]
\item $L = L_{X_0}$, the usual Laplacian for all levels of perturbation.
\item The best performance among $L_{X_i}$ for each level of perturbation.
\item $L = L_{X_b}$ with the best overall performance for all levels of perturbation ($b=2$ in our case).
\item Anomaly is declared when at least $1/3$ of $L_{X_i}, 0\leq i\leq p$ say so. 
\end{enumerate}

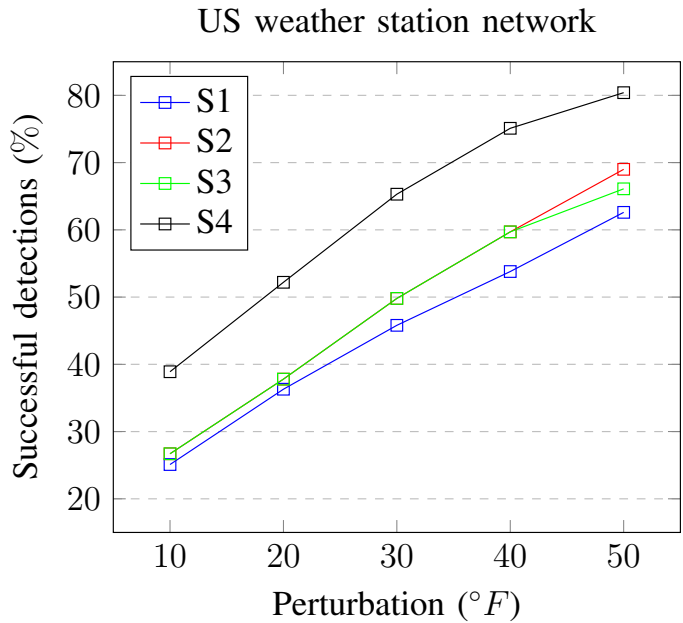
\begin{figure}
	\centering
	\begin{tikzpicture}[scale = 1.1][baseline]

		\begin{scope}[xshift=10cm]
	\begin{axis}[
	title={US weather station network},
	xlabel={Perturbation ($^{\circ}F$)}, 
	ylabel={Successful detections ($\%$)},
	xmin=5, xmax=55,
	ymin=15, ymax=85,
	xtick={10,20,30,40, 50},
	ytick={20, 30, 40, 50, 60, 70, 80},
	legend pos=north west,
	ymajorgrids=true,
	grid style=dashed,
	]
	
	\addplot
	[
	color=blue,
	mark=square,
	]
	coordinates {
		(10,25.1)(20,36.3)(30, 45.8)(40,53.8)(50, 62.6)
	};
	\addlegendentry{S1}
	
	\addplot
	[
	color = red,
	mark = square,
	]
	coordinates{
		(10,26.7)(20,37.8)(30, 49.8)(40,59.7)(50,69.0)
	};
	\addlegendentry{S2}
	
	\addplot
	[
	color = green,
	mark = square,
	]
	coordinates{
		(10,26.7)(20,37.8)(30, 49.8)(40,59.7)(50,66.1)
	} ;
	\addlegendentry{S3}    
	
	\addplot
	[
	color = black,
	mark = square,
	]
	coordinates{
		(10,38.9)(20,52.2)(30, 65.3)(40,75.1)(50,80.4)
	};
	\addlegendentry{S4}    
	
	\end{axis}
	
	\end{scope}
	\end{tikzpicture}
	\caption{Performance of anomaly detection on the US weather station network.} 
	\label{fig:sc7}
\end{figure}

The results are summarized in Figure~\ref{fig:sc7}. We see that in general, we do gain benefits by working with a simplicial complex instead of a graph. The best simplicial structure is $X_2$ when approximately $10\%$ of pair-wisely connected triples are added as $2$-simplexes. It has a consistent overall performance. Moreover, by aggregating observations from different $L_{X_i}$ together, we may further enhance the detection accuracy.

\subsection{Noisy label correction} \label{sec:denoise}

In this section, we consider noisy label correction with the following experimental setup. On a graph $G$ of size $n$, suppose every node $v$ belongs to a class among $k$ classes, and thus has a class label $x(v) \in \{1,\ldots, k\}$. We assume that a certain percentage of the labels are corrupted by noise. The task is to recover the true label as far as possible. 

An approach to the task is to apply a convolution filter to the noisy labels. More specifically, let $x_b$ be the signal of noisy labels and $L$ be a shift operator. Fix a number $0<r<1$ and a scaling factor $0\leq s < 1$. We first find the Fourier transform $\hat{x}_b$ of $x_b$ w.r.t. $L$. To denoise, we scale down $\hat{x}_b(i), rn\leq i\leq n$ by the factor $s$ to obtain $y_b$. To obtain the recovered label, we round off the inverse transform of $y_b$.

The purpose of this paper is to investigate the gain and loss with simplicial complexes over plain graphs. Therefore, we apply the same set of parameters for different choices of $L$. As in Section~\ref{sec:gl} and Section~\ref{sec:ano}, we construct $X_i$ and $L_{X_i}$ for $0\leq i\leq p=10$. We choose a smaller $p$ as we have seen earlier that the spectrum of $L_{X_i}$ may stabilize quickly. 

The graphs we consider are citation graphs: Citeseer ($2120$ nodes, $3679$ edges and $1084$ triangles) and Cora ($2485$ nodes, $5069$ edges and $1558$ triangles) \cite{Sen08, Kip16}. We briefly recall that for both graphs each node represents a document, and the label is the category of the document. The edges are citations links, forming the citation graphs. 

\begin{figure}[!ht]
	\centering
	\includegraphics[height=120mm, width=120mm]{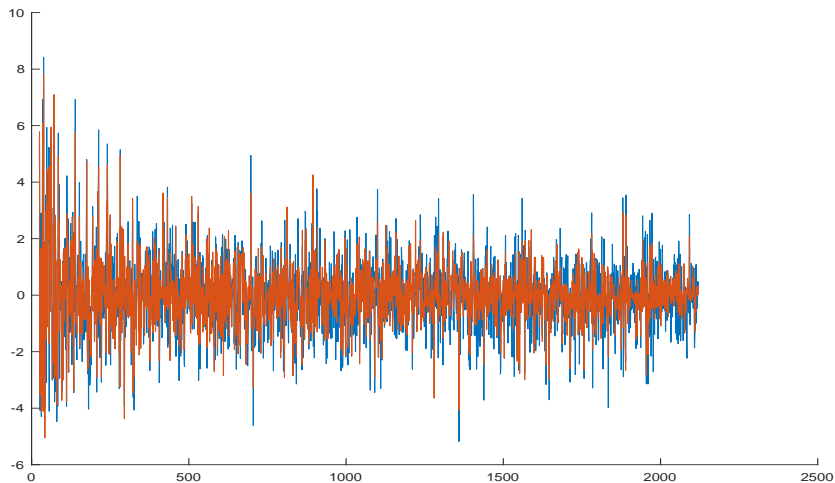}
	\caption{A sample frequency plot of the actual label $x$ (red) and noisy label $x_b$ (blue) in the range $\approx 0.01n$ to $n$. We observe fluctuations for both $x$ and $x_b$, while the amplitudes of $x$ is smaller in general.}\label{fig:sc8}
\end{figure}

For each experiment, we add noise to randomly selected $60\%$ of the labels, with a constant signal-to-noise ratio (SNR). We perform denoising with $r=0.01$ and $s=0.9$, as suggest by sample frequency plot of $x$ and $x_b$ shown in Figure~\ref{fig:sc8}. 

In terms of the reducing the amount of error labels, different $L_{X_i}, 0\leq i\leq p$ perform without noticeable difference. Therefore, for each instance, we determine the $L_{X_i}$ yielding the largest amount of correct labels. Therefore for each $i$, we can estimate the fraction of instances $L_{X_i}$ being the best. The results are summarized in Table~\ref{tab:sc1} and \ref{tab:sc2}. We highlight three $3$ top performers across each row of the tables. 

\begin{table}[!htb]
	\centering  
	\scalebox{1.1}{
		\begin{tabular}{|l|c|c|c|c|c|c|c|c|c|c|c|} 
			\hline
			\emph{SNR} & $L_{X_0}$ & $L_{X_1}$  & $L_{X_2}$ & $L_{X_3}$ & $L_{X_4}$ & $L_{X_5}$ & $L_{X_6}$ & $L_{X_7}$ & $L_{X_8}$ & $L_{X_9}$ & $L_{X_{10}}$  \\ 
			\hline\hline
			$2$ & $\blue{0.127}$ & $\bm{0.126}$  & $0.084$ & $\bm{0.105}$ & $0.103$ & $0.076$ & $0.086$ & $0.103$ & $0.096$ & $0.046$ & $0.046$\\
			\hline
			$1$ & $\bm{0.118}$ & $\blue{0.138}$  & $0.076$ & $0.089$ & $\bm{0.111}$ & $0.088$ & $0.086$ & $0.093$ & $0.100$ & $0.051$ & $0.049$\\
			\hline
			$0$ & $\bm{0.118}$ & $\blue{0.131}$  & $0.079$ & $0.094$ & $\bm{0.114}$ & $0.087$ & $0.101$ & $0.099$ & $0.094$ & $0.041$ & $0.041$\\
			\hline
			$-1$ & $\bm{0.105}$ & $\blue{0.138}$  & $0.093$ & $0.085$ & $0.090$ & $0.093$ & $0.100$ & $\bm{0.110}$ & $0.098$ & $0.044$ & $0.044$\\
			\hline
			$-2$ & $\bm{0.127}$ & $\blue{0.136}$  & $0.083$ & $0.093$ & $0.082$ & $0.092$ & $0.094$ & $0.096$ & $\bm{0.102}$ & $0.047$ & $0.048$\\
			\hline
	\end{tabular}}
	\caption{Citeseer} \label{tab:sc1}
\end{table}    

\begin{table}[!htb]
	\centering  
	\scalebox{1.1}{
		\begin{tabular}{|l|c|c|c|c|c|c|c|c|c|c|c|} 
			\hline
			\emph{SNR} & $L_{X_0}$ & $L_{X_1}$  & $L_{X_2}$ & $L_{X_3}$ & $L_{X_4}$ & $L_{X_5}$ & $L_{X_6}$ & $L_{X_7}$ & $L_{X_8}$ & $L_{X_9}$ & $L_{X_{10}}$  \\ 
			\hline\hline
			$2$ & $\bm{0.134}$ & $\blue{0.172}$  & $0.099$ & $\bm{0.121}$ & $0.083$ & $0.084$ & $0.050$ & $0.087$ & $0.060$ & $0.051$ & $0.060$\\
			\hline
			$1$ & $\bm{0.143}$ & $\blue{0.163}$  & $0.063$ & $\bm{0.126}$ & $0.125$ & $0.063$ & $0.083$ & $0.094$ & $0.053$ & $0.040$ & $0.048$\\
			\hline
			$0$ & $\bm{0.115}$ & $\blue{0.133}$  & $0.101$ & $0.085$ & $0.103$ & $\bm{0.129}$ & $0.077$ & $0.041$ & $0.051$ & $0.070$ & $0.091$\\
			\hline
			$-1$ & $\blue{0.136}$ & $\bm{0.134}$  & $0.085$ & $\bm{0.127}$ & $0.049$ & $0.093$ & $0.088$ & $0.056$ & $0.046$ & $0.066$ & $0.120$\\
			\hline
			$-2$ & $\bm{0.124}$ & $0.093$  & $\bm{0.132}$ & $\blue{0.153}$ & $0.089$ & $0.078$ & $0.042$ & $0.059$ & $0.076$ & $0.089$ & $0.065$\\
			\hline
	\end{tabular}}
	\caption{Cora} \label{tab:sc2}
\end{table}    

From the results, we observe that $L_{X_1}$ has the best overall performance consistently, which is when approximately $10\%$ of pair-wisely connected triples are added as $2$-simplexes. As a general trend, the perform drops if a large amount of $2$-simplexes are added as in $X_9$ and $X_{10}$. We do gain benefits by working with high dimensional components against the plain graphs.

\section{Conclusions} \label{sec:con}
In this paper, we have proposed a signal processing framework for signals on simplicial complexes. To do so, we introduced a general way to construct a Laplacian matrix on a space, which may not be a graph. After which, signal processing follows in the way similar to traditional GSP. We test the framework with both synthetic and real datasets, and observe that we do gain benefits by working with additional high structures. 

A lot of new signal processing techniques and problems may stem from our new framework. For future works, we shall exploring such possibilities including continuous filter estimation and data driven based end-to-end learning.

\appendices

\section{Non shift invariance} \label{sec:nsi}

In this appendix, we assume $X$ is a $2$-complex of size $n$ and discuss conditions that ensure $L_X$ is not shift invariant w.r.t.\ $L_{X^1}$. We are mainly interested in geometric conditions, which can be observed directly from the shape of $X$. As a corollary, we prove Proposition~\ref{prop:sxi}. Most of the notations follow those defined in Section~\ref{sec:2cs}.

For convenience, we introduce the following notion. 

\begin{Definition}
	If a matrix $M$ is the Laplacian of a weighted graph $G$, then we say $M$ is of graph type $G$. Moreover, we say that $X$ has \emph{distinctive $2$-simplexes} if (a) either $L_X-L_{X^1}$ or $L_{X^1}-L_X$ is of graph type $G$; and (b) an edge $e=(v_1,v_2)$ of $G$ has positive edge weights when $e$ belongs to a $2$-simplex of $X$.
\end{Definition}

\begin{Lemma}
	$X$ has distinctive $2$-simplexes if either of the following holds:
	\begin{enumerate}[(a)]
		\item $k_{\max} \leq 1$, i.e., each edge is contained in at most one $2$-simplex.
		
		\item $k_{\max} \leq 2$ and all the edges have weight $1$.
		
		\item $k_{\min} \geq 4$ and all the edges have weight $1$.
	\end{enumerate}
\end{Lemma}

\begin{IEEEproof}
	(a) As any constant vector is in the kernel of $L = L_{X^1}-L_X$, the sum of each row is $0$. If $(i,j)$ is an edge of $X$ not contained in any $2$-simplex, then the $(i,j)$-th entry of $L$ is $0$. It suffices to show that if $(v_i,v_j)$ is any edge contained in a $2$-simplex, then the $(i,j)$-th entry of $L$ is negative. Let $a>0$ be the weight of $(v_i,v_j)$ and $b>0, c>0$ be the weights of the other two edges of the $2$-simplex containing $(v_i,v_j)$. A direct calculation shows that the $(i,j)$-th entry of $L$ is $-(13a+b+c)/18 <0$.
	
	(b) and (c) can be shown by the same argument by considering $L_{X^1}-L_X$ and $L_X-L_{X^1}$ respectively. 
\end{IEEEproof}

Assume for the rest of this appendix that $X$ has distinctive $2$-simplexes. We want to study common eigenvectors of both $L_{X^1}$ and $L_X$. To this end, we divide the discussion into two parts: for such an eigenvector, whether the associated eigenvalues are the same or different.

\begin{figure}[!ht]
	\centering
	\includegraphics[scale=2.5]{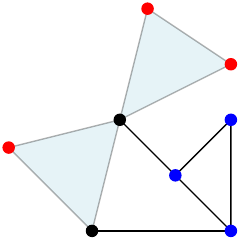}
	\caption{In this example, all the blue nodes are $1$-interior and red nodes are $2$-interior. Hence, for the parameters in Lemma~\ref{lem:tkk}\ref{it:lmb} $m_1=3$ and $m_2=m_3=1$. Moreover, in Lemma~\ref{lem:swi}\ref{it:ivb}, $m_4=2$ counts the two black nodes.} \label{fig:sc4}
\end{figure}

\begin{Definition}
We call a vertex $v$ is $1$-interior if it is not contained in any $2$-simplex and $2$-interior if each edge containing $v$ belongs to a $2$-simplex (see Figure~\ref{fig:sc4} for an example).
\end{Definition}

\begin{Lemma} \label{lem:tkk}
	\begin{enumerate}[(a)]
		\item Let $K$ be the vector space spanned by common eigenvectors with the same eigenvalue of $L_X$ and $L_{X^1}$. Then $K$ is a subspace of $\ker(L_X-L_{X^1})$.
		
		\item \label{it:lmb} Let $m_1$ be the number of $1$-interior nodes of $X$, $m_2$ be the number of connected components of smallest complex containing all the $2$-simplexes of $X$, and $m_3$ be the number of such components containing some $2$-interior nodes. Then $\dim K \leq m_1+m_2-m_3$.
	\end{enumerate}
\end{Lemma}

\begin{IEEEproof}
	(a) As we assume that $X$ has distinctive $2$-simplexes, $L_{X^1}-L_X = L_G$ or $-L_G$ for some graph $G$ whose positive edge weights are supported on $2$-simplexes of $X$. Therefore, if $w$ is a common eigenvector with the same eigenvalue, then $L_G(x)=0$, i.e., $w\in \ker(L_G)$. As $\ker(L_G)$ is a vector space, $K$ as spanned by these $v$'s is also contained in $\ker(L_G)$.
	
	(b) Notice that $\ker(L_G)$ is the same as the number of connected components of $G$. The set of connected components of $G$ consists of: (1) each $1$-interior node of $X$ is an isolated component of $G$, (2)  a union of $2$-simplexes that is connected. They are of size $m_1$ and $m_2$ respectively. Suppose a component $C$ of the second type contains a $2$-interior node $v$ and $w$ is a common eigenvector with the same eigenvalue $\lambda>0$. Then $L_{X^1}(w)(v) = L_G(w)(v) = 0$. However, $L_{X^1}(w)(v) =\lambda w(v)$, and hence $w(v)=0$. Hence, $w$ is $0$ are all of $C$ as $w$ is constant on $C$. Hence, the vectors of $K$ vanishes on such a $C$. Therefore, $\dim K \leq m_1+m_2-m_3$.   
\end{IEEEproof} 

Now we consider common eigenvectors of $L_{X^1}$ and $L_X$ with different eigenvalues. 

\begin{Lemma} \label{lem:swi}
	Suppose $w$ is a common eigenvectors of $L_{X^1}$ and $L_X$ with different eigenvalues. Then
	\begin{enumerate}[(a)]
	\item $w$ is $0$ at $1$-interior nodes of $X$.
	
	\item \label{it:ivb} If $v$ belongs to a $2$-simplex and any $1$-interior neighbor of $v$ is not connected to any other node belonging to a $2$-simplex, then $w$ is $0$ at $v$. Denote the number of such vertices by $m_4$.
	\end{enumerate}
\end{Lemma}

\begin{IEEEproof}
	Suppose the eigenvalues of $w$ are $\lambda_1 \neq \lambda_2$.
	
	(a) Let $v$ be a $1$-interior node. Then $L_X(w)(v) = L_{X^1}(w)(v)$ as the neighborhood of $v$ in $X$ and $X^1$ are identical. This implies that $\lambda_1 w(v) = \lambda_2 w(v)$. This is possible only if $w(v)=0$.
	
	(b) Let $v'$ be a $1$-interior neighbor of $v$. By (a), $w(v')=0$. As $v'$ is not connected to any other node belonging to a $2$-simplex, $w$ is $0$ at the neighbors of $v'$ except at $v$. Hence $0 = L_{X^1}(w)(v') = a w(v)$, where $a$ is the positive edge weight between $v$ and $v'$. This proves (b).
\end{IEEEproof}

Now, we are ready to state and prove the main result of this section.

\begin{Theorem} \label{thm:idt}
	If $\dim K \leq m_1+m_4 < n$, then there does not exist any orthonormal basis consisting of common eigenvectors of both $L_{X^1}$ and $L_X$. In particular, this holds if $m_2 \leq m_3+m_4$.
\end{Theorem}

\begin{IEEEproof}
	Suppose on the contrary that $W = \{w_1,\ldots, w_n\}$ (column vectors) is an orthonormal basis consisting of common eigenvectors of $L_{X^1}$ and $L_X$. There are at most $\dim K$ vectors of $W$ each shares the same eigenvalue. Without loss of generality, assume they are $\{w_1,\ldots, w_{\dim K}\}$ and let $w_1$ be the constant vector $(1/\sqrt{n},\ldots,1/\sqrt{n})'$. Moreover, by re-indexing, we further assume that the first $m_1+m_4$ indices correspond to the set $S$ of $1$-interior nodes and nodes satisfy Lemma~\ref{lem:swi}\ref{it:ivb}. 
	
	By abuse of notation, write $W$ for the $n\times n$ matrix whose $i$-th column being $w_i$. As the columns of $W$ forms a orthonormal basis, so do the rows of $W$. On the other hand, by Lemma~\ref{lem:swi}, only the leading $(m_1+m_4)\times \dim K$ block $W_1$ of the first $m_1+m_4$ rows of $W$ can contain non-zero entries. Hence, the rows of $W_1$ forms an orthonormal system. This shows that $m_1+m_4 \leq \dim K$.
	
	We claim $m_1+m_4 \neq \dim K$. For otherwise, $W_S$ is a $\dim K\times \dim K$ matrix with orthonormal rows. Hence, the columns of $W_S$ also forms an orthonormal system. However, this is impossible as the norm of the first column of $W_S$ is $\dim K/n <1$.
	
	Therefore, we have shown that $m_1+m_4 < \dim K$ with the existence of $W$. This contradicts the assumption that $\dim K \leq m_1+m_4$. Furthermore, the condition $m_2 \leq m_3+m_4$ implies that $\dim K \leq m_1+m_4$ by Lemma~\ref{lem:tkk}\ref{it:lmb}.
\end{IEEEproof}

As a corollary, we can prove Proposition~\ref{prop:sxi} by counting. First of all, by Condition~\ref{it:edi}, $X$ has distinctive $2$-simplexes. In order to show $L_X$ is not shift invariant w.r.t.\ $L_{X^1}$, we want to prove that they cannot have a common orthonormal eigenbasis. By Theorem~\ref{thm:idt}, it suffices to show that $m_2\leq m_4$ under the assumptions of Theorem~\ref{thm:idt}. Let $C$ be a union of $2$-simplexes contributing $1$ to $m_2$ in $X$. In $C$, there is at least one vertex $v_C$ connected to a $1$-interior point for otherwise, we can either add another $2$-simplex to enlarge $C$ or $X$ contains no $1$-interior point, which contradicts Condition~\ref{it:ixi}. Moreover, $v_C$ cannot be shared by another connected union of $2$-simplexes by Condition~\ref{it:ixa}. In conclusion, $C\mapsto v_C$ is a one-one map and hence $m_2\leq m_4$, and Proposition~\ref{prop:sxi} follows.

\bibliographystyle{IEEEtran}
\bibliography{IEEEabrv,StringDefinitions,biblio}

\begin{thebibliography}{10}
\providecommand{\url}[1]{#1}
\csname url@samestyle\endcsname
\providecommand{\newblock}{\relax}
\providecommand{\bibinfo}[2]{#2}
\providecommand{\BIBentrySTDinterwordspacing}{\spaceskip=0pt\relax}
\providecommand{\BIBentryALTinterwordstretchfactor}{4}
\providecommand{\BIBentryALTinterwordspacing}{\spaceskip=\fontdimen2\font plus
\BIBentryALTinterwordstretchfactor\fontdimen3\font minus
  \fontdimen4\font\relax}
\providecommand{\BIBforeignlanguage}[2]{{%
\expandafter\ifx\csname l@#1\endcsname\relax
\typeout{** WARNING: IEEEtran.bst: No hyphenation pattern has been}%
\typeout{** loaded for the language `#1'. Using the pattern for}%
\typeout{** the default language instead.}%
\else
\language=\csname l@#1\endcsname
\fi
#2}}
\providecommand{\BIBdecl}{\relax}
\BIBdecl

\bibitem{Shu13}
D.~I. Shuman, S.~K. Narang, P.~Frossard, A.~Ortega, and P.~Vandergheynst, ``The
  emerging field of signal processing on graphs: Extending high-dimensional
  data analysis to networks and other irregular domains,'' \emph{IEEE Signal
  Process. Mag.}, vol.~30, no.~3, pp. 83--98, May 2013.

\bibitem{San13}
A.~Sandryhaila and J.~M.~F. Moura, ``Discrete signal processing on graphs,''
  \emph{IEEE Trans. Signal Process.}, vol.~61, no.~7, pp. 1644--1656, April
  2013.

\bibitem{San14}
------, ``Big data analysis with signal processing on graphs: Representation
  and processing of massive data sets with irregular structure,'' \emph{IEEE
  Signal Process. Mag.}, vol.~31, no.~5, pp. 80--90, Sept 2014.

\bibitem{Don16}
X.~Dong, D.~Thanou, P.~Frossard, and P.~Vandergheynst, ``Learning laplacian
  matrix in smooth graph signal representations,'' \emph{IEEE Transactions on
  Signal Processing}, vol.~64, no.~23, pp. 6160--6173, 2016.

\bibitem{ort17}
H.~E. Egilmez, E.~Pavez, and A.~Ortega, ``Graph learning from data under
  laplacian and structural constraints,'' \emph{IEEE Journal of Selected Topics
  in Signal Processing}, vol.~11, no.~6, pp. 825--841, 2017.

\bibitem{Gra18}
F.~Grassi, A.~Loukas, N.~Perraudin, and B.~Ricaud, ``A time-vertex signal
  processing framework: Scalable processing and meaningful representations for
  time-series on graphs,'' \emph{IEEE Trans. Signal Process.}, vol.~66, no.~3,
  pp. 817--829, Feb 2018.

\bibitem{Ort18}
A.~Ortega, P.~Frossard, J.~Kova{\v{c}}evi{\'c}, J.~M. Moura, and
  P.~Vandergheynst, ``Graph signal processing: Overview, challenges, and
  applications,'' \emph{Proceedings of the IEEE}, vol. 106, no.~5, pp.
  808--828, 2018.

\bibitem{JiTay:J19}
F.~Ji and W.~P. Tay, ``A {Hilbert} space theory of generalized graph signal
  processing,'' \emph{{IEEE} Trans. Signal Process.}, vol.~67, no.~24, pp. 6188
  -- 6203, Dec. 2019.

\bibitem{Def16}
M.~Defferrard, X.~Bresson, and P.~Vandergheynst, ``Convolutional neural
  networks on graphs with fast localized spectral filtering,'' in
  \emph{Advances in Neural Inform. Process. Syst.}, USA, 2016, pp. 3844--3852.

\bibitem{Kip16}
T.~N. Kipf and M.~Welling, ``Semi-supervised classification with graph
  convolutional networks,'' \emph{arXiv preprint arXiv:1609.02907}, 2016.

\bibitem{Wan18}
R.~Li, S.~Wang, F.~Zhu, and J.~Huang, ``Adaptive graph convolutional neural
  networks,'' in \emph{Thirty-second AAAI conference on artificial
  intelligence}, 2018.

\bibitem{klamt2009hypergraphs}
S.~Klamt, U.-U. Haus, and F.~Theis, ``Hypergraphs and cellular networks,''
  \emph{PLoS computational biology}, vol.~5, no.~5, p. e1000385, 2009.

\bibitem{flamm2015generalized}
C.~Flamm, B.~M. Stadler, and P.~F. Stadler, ``Generalized topologies:
  hypergraphs, chemical reactions, and biological evolution,'' in
  \emph{Advances in Mathematical Chemistry and Applications}.\hskip 1em plus
  0.5em minus 0.4em\relax Elsevier, 2015, pp. 300--328.

\bibitem{lohmann2012representing}
S.~Lohmann and P.~D{\'\i}az, ``Representing and visualizing folksonomies as
  graphs: a reference model,'' in \emph{Proceedings of the International
  Working Conference on Advanced Visual Interfaces}, 2012, pp. 729--732.

\bibitem{Alt92}
N.~Altman, ``An introduction to kernel and nearest-neighbor nonparametric
  regression.''

\bibitem{Don19}
X.~{Dong}, D.~{Thanou}, M.~{Rabbat}, and P.~{Frossard}, ``Learning graphs from
  data: A signal representation perspective,'' \emph{IEEE Signal Process.
  Mag.}, vol.~36, no.~3, pp. 44--63, 2019.

\bibitem{Mat19}
G.~{Mateos}, S.~{Segarra}, A.~G. {Marques}, and A.~{Ribeiro}, ``Connecting the
  dots: Identifying network structure via graph signal processing,'' \emph{IEEE
  Signal Process. Mag.}, vol.~36, no.~3, pp. 16--43, 2019.

\bibitem{Ram19}
M.~{Ramezani-Mayiami}, M.~{Hajimirsadeghi}, K.~{Skretting}, R.~S. {Blum}, and
  H.~{Vincent Poor}, ``Graph topology learning and signal recovery via bayesian
  inference,'' in \emph{2019 IEEE Data Science Workshop (DSW)}, 2019, pp.
  52--56.

\bibitem{Ji20}
F.~Ji, W.~Tang, W.~P. Tay, and E.~K.~P. Chong, ``Network topology inference
  using information cascades with limited statistical knowledge,''
  \emph{Information and Inference: A Journal of the {IMA}}, 2020.

\bibitem{Spa66}
E.~H. Spanier, \emph{Algebraic topology}.\hskip 1em plus 0.5em minus
  0.4em\relax Springer Science \& Business Media, 1989.

\bibitem{carlsson2009topology}
G.~Carlsson, ``Topology and data,'' \emph{Bulletin of the American Mathematical
  Society}, vol.~46, no.~2, pp. 255--308, 2009.

\bibitem{Cou16}
O.~T. Courtney and G.~Bianconi, ``Generalized network structures: The
  configuration model and the canonical ensemble of simplicial complexes,''
  \emph{Phys. Rev. E}, vol.~93, p. 062311, Jun 2016.

\bibitem{bar09}
S.~Barbarossa and S.~Sardellitti, ``Topological signal processing over
  simplicial complexes,'' \emph{arXiv preprint arXiv:1907.11577}, 2019.

\bibitem{bar16}
S.~Barbarossa and M.~Tsitsvero, ``An introduction to hypergraph signal
  processing,'' in \emph{EEE Int. Conf. Acoustics, Speech and Signal Process.},
  2016, pp. 6425--6429.

\bibitem{pus19}
M.~Puschel, ``A discrete signal processing framework for meet/join lattices
  with applications to hypergraphs and trees,'' in \emph{IEEE Int. Conf.
  Acoustics, Speech and Signal Process.}, 2019, pp. 5371--5375.

\bibitem{zha19}
S.~Zhang, Z.~Ding, and S.~Cui, ``Introducing hypergraph signal processing:
  theoretical foundation and practical applications,'' \emph{arXiv preprint
  arXiv:1907.09203}, 2019.

\bibitem{Hat02}
A.~Hatcher, \emph{Algebraic Topology}.\hskip 1em plus 0.5em minus 0.4em\relax
  Cambridge University Press, 2002.

\bibitem{Ber73}
C.~Berge, \emph{Graphs and Hypergraphs}.\hskip 1em plus 0.5em minus 0.4em\relax
  American Elsevier Pub. Co, 1973.

\bibitem{Ouv20}
X.~Ouvrard, ``Hypergraphs: an introduction and review,''
  \emph{arXiv:2002.05014}, 2020.

\bibitem{Kap02}
I.~Kapovich and N.~Benakli, ``Boundaries of hyperbolic groups,'' \emph{arXiv
  preprint math/0202286}, 2002.

\bibitem{JiT19}
F.~Ji, W.~Tang, and W.~P. Tay, ``On the properties of {Gromov} matrices and
  their applications in network inference,'' \emph{{IEEE} Trans. Signal
  Process.}, vol.~67, no.~10, pp. 2624 -- 2638, May 2019.

\bibitem{Kli04}
B.~Klimt and Y.~Yang, ``Introducing the enron corpus.'' in \emph{CEAS}, 2004.

\bibitem{Sen08}
P.~Sen, G.~Namata, M.~Bilgic, L.~Getoor, B.~Galligher, and T.~Eliassi-Rad,
  ``Collective classification in network data,'' \emph{AI mag.}, vol.~29,
  no.~3, p.~93, 2008.

\end{thebibliography}

\end{document}